\documentclass[reprint,prd,twocolumn,superscriptaddress,nofootinbib,floatfix]{revtex4-1}
\usepackage{graphicx, bm,color,amsmath,amssymb}
\newcommand{\lsim}{\mbox{\raisebox{-.6ex}{~$\stackrel{<}{\sim}$~}}}

\begin{document}
\title{Probing the Scale of New Physics by Advanced LIGO/VIRGO}

\author{P. S. Bhupal Dev}
\affiliation{Max-Planck-Institut f\"{u}r Kernphysik, Saupfercheckweg 1, D-69117 Heidelberg, Germany}

\author{A. Mazumdar}
\affiliation{Consortium for Fundamental Physics, Lancaster University, Lancaster, LA1 4YB, United Kingdom}

\begin{abstract}
We show that if the new physics beyond the Standard Model is associated with a first-order phase transition around $10^7$--$10^8$ GeV, the energy density stored in the resulting stochastic gravitational waves and the corresponding peak frequency are within the projected final sensitivity of the advanced LIGO/VIRGO detectors. 
 We discuss some possible new physics scenarios that could arise at such energies, and in particular, the consequences for Peccei-Quinn and supersymmetry breaking scales.
\end{abstract}
\date{\today}
\maketitle

\section{Introduction}
Recently, the two detectors of the advanced Laser Interferometer Gravitational-Wave Observatory (LIGO) observed a transient gravitational-wave (GW) signal with a significance in excess of 5.1$\sigma$~\cite{Abbott:2016blz}. This spectacular signal is consistent with a binary black hole merger with initial black hole masses of $36^{+5}_{-4} M_\odot$ and $29^{+4}_{-4} M_\odot$, and the final black hole mass of $62^{+4}_{-4} M_\odot$, as measured in the source frame, at a luminosity distance of $410^{+160}_{-180}$ Mpc corresponding to a redshift of $z = 0.09^{+0.03}_{-0.04}$. This event, named GW150914, inaugurates a new era of GW astronomy, as we begin ``hearing" from the Universe. 

In fact, there are several known sources giving rise to potentially observable gravitational waves, which can be broadly split into three categories~\cite{Maggiore:1999vm}: (i) transient signals emitted by binary black hole mergers, coalescing binary neutron stars or a neutron star and a black hole, or supernova core collapse, with a duration between a millisecond and several hours; (ii) long-duration signals, e.g. from spinning neutron stars; and (iii) stochastic background arising from the superposition of unresolved astrophysical sources. A stochastic background of gravitational waves can also arise from cosmological events, such as during primordial inflation~\cite{inflation}, after inflation during resonant preheating~\cite{preheating}, or due to fragmentation of the inflaton or any scalar condensate~\cite{Kusenko:2008zm}, cosmic strings~\cite{Damour,Aasi:2013vna}, and cosmological phase transitions~\cite{Kosowsky:1992rz,Kamionkowski:1993fg},  which can be potentially constrained by the current and future GW experiments, since our Universe is transparent to gravitational waves all the way back to the Planck epoch. This provides an unprecedented opportunity to study some of these cosmological phenomena never seen before. 

Motivated by the LIGO discovery, we study the possibility of observing the stochastic GW background from a strong {\it first-order} cosmological phase transition in future data. First-order phase transitions are predicted in many scenarios beyond the Standard Model (BSM), including one which is associated with the electroweak (EW) scale that might be responsible for the observed baryon asymmetry of our Universe. It is known~\cite{Kosowsky:1992rz,Kamionkowski:1993fg,Grojean:2006bp,Caprini:2007xq, Huber:2008hg,Espinosa:2010hh,
Schwaller:2015tja,Kakizaki:2015wua, Caprini:2015zlo,Huang:2016odd} that the peak frequency of the gravitational waves produced in the first-order phase transition with a critical temperature close to the EW scale is around mHz, which is much below the frequency range (10--500 Hz) of ground-based GW detectors, and can only be probed in future space-based GW experiments, such as eLISA~\cite{AmaroSeoane:2012km}. However, we show that if similar first-order phase transitions occurred at a critical temperature of 
${\cal O}$($10^7$--$10^8$) GeV, this could potentially give rise to a GW signal in the observable range of advanced LIGO/VIRGO~\cite{TheLIGOScientific:2014jea, TheVirgo:2014hva}, and provide us with a unique probe of the BSM physics at such high scales not directly accessible by laboratory means. We will discuss a few well-motivated BSM examples in this context, such as the Peccei-Quinn (PQ) symmetry breaking~\cite{Peccei:1977hh} in order to solve the strong CP problem by an axion~\cite{Weinberg:1977ma} and high-scale Supersymmetry (SUSY) breaking within the framework of some well-known SUSY models, e.g. Next-to-Minimal Supersymmetric Standard Model (NMSSM)~\cite{Ellwanger:2009dp}, split-SUSY~\cite{Giudice:2004tc}, and SUSY breaking mediated by sneutrinos~\cite{Mohapatra:1986uf}. 

Before proceeding further, we would like to make a general comment that the gravitational waves are typically assumed to be quantum in nature, where the force carriers, {\it viz.}  gravitons, are spin-2 quanta described by the linearized quantization of the Einstein-Hilbert action~\cite{Maggiore:1999vm}. The LIGO detection as such does not confirm whether the observed gravitational wave is classical or quantum. A recent proposal is to search for anomalies such as decreased regularity of the signal and increased power~\cite{Giddings:2016tla}. On the other hand, any positive detection of primordial B-modes in the cosmic microwave background (CMB) radiation would be a clear evidence for the quantum nature of the gravity waves~\cite{Ashoorioon:2012kh}, see also~\cite{Krauss:2013pha}. 

The rest of the paper is organized as follows: In Section~\ref{sec:2}, we briefly discuss the essence of first-order phase transition. 
In Section~\ref{sec:3}, we calculate the spectrum of gravitational waves from phase transition and make a comparison with the astrophysical and inflationary GW signals for a possible distinction in future data. In Section~\ref{sec:4}, we discuss a few BSM scenarios which could give rise to an observable GW signal through a strong first-order phase transition at high scale. Our conclusion is given in Section~\ref{sec:5}.
\section{First-order Phase Transition}  \label{sec:2}  
Typically, a first-order cosmological phase transition can occur when two local minima of the free energy co-exist, during which our Universe could be realizable in the metastable vacuum, or a false-vacuum state for some range of temperatures. The transition to the true vacuum state is achieved only by quantum-mechanical tunneling or by thermal fluctuations~\cite{Callan:1977pt, Kolb:1988aj}. If the energy barrier between the false and true vacuum states is sufficiently large, these quantum or thermal processes proceed through the nucleation and percolation of bubbles  of true vacuum in a sea of metastable phase.  Once nucleated, the bubble expands outward with constant acceleration driven by the pressure difference between its true-vacuum interior and false-vacuum exterior, and quickly approaches the speed of light, unless there is significant friction due to coupling with the thermal bath. The bubbles will eventually collide and a large amount of the false-vacuum energy existing in the form of the bubble-wall kinetic energy is dumped into the ambient thermal bath of radiation. This sequence of events can give rise to an observable stochastic GW background in the LIGO sensitivity range for some critical temperatures, as we show in Section~\ref{sec:3}. 

The bubble nucleation rate per unit volume is given by $\Gamma(t) =\Gamma_0(t) e^{-S(t)}$, where $S$ is the 4-dimensional Euclidean action of a critical bubble~\cite{Kolb:1988aj}. The inverse time duration of the phase transition is given by 
\begin{align}
\beta \ \equiv \ -\left.\frac{dS}{dt}\right|_{t=t_*} \ \simeq \ \left.\frac{d\ln \Gamma}{dt}\right|_{t=t_*},
\end{align}
where $t_*$ denotes the time when gravitational waves are produced. A key parameter controlling the energy density of the GW signal is the fraction $\beta/H_*$, where 
\begin{align}
H_*(t) \ = \ \left[\frac{8\pi^3g_*(t)T^4_*(t)}{90 \: m_{\rm Pl}^2}\right]^{1/2}
\end{align}
 is the Hubble parameter and $g_*$ is the number of relativistic degrees of freedom (d.o.f) in the thermal plasma, both evaluated at the critical temperature $T_*$ (which is approximately equivalent to the nucleation temperature for typical phase transitions without significant reheating), and $m_{\rm Pl}=1.22\times 10^{19}$ GeV is the Planck mass. 

Another key parameter measuring the strength of the phase transition is the ratio of the false vacuum energy density released in the process to that of the ambient plasma thermal energy density at $T_*$; denoted by 
\begin{align}
\alpha \ \equiv \ \frac{\rho_{\rm vac}}{\rho_*} \, , 
\end{align}
where $\rho_*=g_*\pi^2T_*^4/30$ in the symmetric phase~\cite{Kolb:1988aj}. 


With these definitions, the fraction of energy liberated into gravitational waves by collisions of bubble walls can be computed analytically by the ``envelope approximation"~\cite{Kamionkowski:1993fg,Caprini:2007xq, Huber:2008hg, Espinosa:2010hh}
\begin{align}
\frac{E_{\rm GW}}{E_{\rm tot}} \ \propto \ \kappa^2 v^3 \left(\frac{\alpha}{1+\alpha}\right)^2 \left(\frac{H_*}{\beta}\right)^2, 
\label{gw1}
\end{align} 
where $v$ is the bubble-wall velocity and $\kappa$ is the efficiency factor quantifying the fraction of the available vacuum energy going into kinetic energy. In a strong first-order phase transition limit, $v\to 1$ and $\alpha \gg 1$, which leads to run-away bubbles in vacuum with $\kappa\to 1$~\cite{Espinosa:2010hh}. In this limit, the strength of the GW signal in Eq.~\eqref{gw1} only depends on the fraction 
\begin{align}
\frac{\beta}{H_*} \ \sim \ \ln\left(\frac{m_{\rm Pl}}{T_*}\right) \, ,
\label{betaH}
\end{align} 
up to a factor of order ${\cal O}(1)$~\cite{Kosowsky:1992rz}. The exact time scale of the phase transition $\beta^{-1}$ is rather difficult to compute in general and one may have to resort to lattice simulations for a better estimate than that given in Eq.~\eqref{betaH}. In concrete models with a given temperature-dependent effective potential, one typically finds~\cite{Jaeckel:2016jlh} 
\begin{eqnarray}
\frac{\beta}{H_\ast} \ = \ T\frac{d}{dT}\left(\frac{S_3}{T}\right)_{T=T_\ast} \ \simeq \ \frac{5}{\epsilon} \, ,
\label{betaH2}
\end{eqnarray}
where $S_3$ is the 3-dimensional spherically-symmetric effective action, and $\epsilon$ is the split
in the energy density between the two vacua. In the thin-wall limit, $0\leq \epsilon\ll 1$ and one usually gets $\beta/H_\ast \sim {\cal O}(100-1000)$ from Eq.~\eqref{betaH2}. However, there exist extreme scenarios where smaller $\beta/H_\ast \sim {\cal O}(1-10)$ is also possible~\cite{Schwaller:2015tja}. 

We should also mention here that, apart from the bubble-wall collisions, there are other potential sources of gravitational waves associated with first-order phase transitions, such as sound waves from the bulk motion in the fluid caused by percolation of bubbles~\cite{Hindmarsh:2013xza} and magnetohydrodynamic turbulences in the plasma after the bubbles have collided~\cite{Caprini:2006jb}. We will not discuss the acoustic and turbulent GW production, since these plasma contributions are not significant in the simple scenario under consideration, namely, run-away bubbles due to phase transitions in a vacuum-dominated epoch~\cite{Caprini:2015zlo}. 

\section{Observable Gravity Waves}\label{sec:3}

To translate Eq.~\eqref{gw1} into a potentially observable GW signal today, we must take into account the redshift factor from the epoch of phase transition, $t_*$, to today, $t_0$. Since the gravitational waves are essentially decoupled from the rest of the Universe, the energy density in gravitational waves simply decreases as $R^{-4}$ and the frequency 
redshifts as $R^{-1}$, where $R$ is the scale factor of the expansion of the Universe. Assuming an adiabatic expansion of the Universe since the phase transition epoch, the ratio of the scale factors is then given by 
\begin{align}
\frac{R_*}{R_0} \ \simeq \ (8.0\times 10^{-14})\left(\frac{100}{g_*}\right)^{1/3}\left(\frac{1~{\rm GeV}}{T_*}\right)\,.
\end{align}
From numerical simulations using the envelope approximation, the peak frequency and the peak value of the fraction of the total energy density in the gravitational waves today are respectively found to be~\cite{Huber:2008hg}
\begin{align}
& f_0 \  \simeq \ (1.65\times 10^{-7}~{\rm Hz})\left(\frac{f_*}{\beta}\right)\left(\frac{\beta}{H_*}\right) \left(\frac{T_*}{1~{\rm GeV}}\right)\left(\frac{g_*}{100}\right)^{1/6}, \label{eq:freq} \\ 
& \Omega_{0}h^2 \ \simeq \ (1.67\times 10^{-5})\kappa^2 \left(\frac{\alpha}{1+\alpha}\right)^2\left(\frac{0.11 \: v^3}{0.42+v^2}\right) \nonumber \\
& \qquad \qquad \qquad \times \left(\frac{H_*}{\beta}\right)^2\left(\frac{100}{g_*}\right)^{1/3},
\label{eq:om}
\end{align}
where $h=0.678\pm 0.009$ is the current value of the Hubble parameter in units of 100 km sec$^{-1}$Mpc$^{-1}$~\cite{Ade:2015xua} and $f_*$ is the peak frequency at $t_*$~\cite{Huber:2008hg}: 
\begin{align}
\frac{f_*}{\beta} \ = \ \frac{0.62}{1.8-0.1\: v+v^2} \, .
\end{align}
 Note that in the strong first-order phase transition limit of $v\to 1$ and $\alpha\gg 1$, Eq.~\eqref{eq:om} reduces to the thin-wall approximation given in Ref.~\cite{Kosowsky:1992rz}, which will be assumed here to be the case for simplicity. 

For a generic first-order phase transition, the spectrum of GW radiation, i.e. energy density per logarithmic frequency interval, normalized to the critical energy density of the Universe, increases as $f^{2.8}$ at low frequencies~\cite{Kosowsky:1992rz} and decreases as $f^{-1}$ at high frequencies~\cite{Huber:2008hg}. These qualitative features can be captured well by a simple parametrization of the spectral shape given by
\begin{align}
\Omega_{\rm GW}(f)h^2 \ = \ \Omega_{0}h^2 \: \frac{(p+q)\left(\frac{f}{f_0}\right)^{p}}{q+p \left(\frac{f}{f_0}\right)^{p+q}} \, , 
\label{eq:gw2}
\end{align}   
with $p=2.8$ and $q=1.0$ from a fit to simulation data~\cite{Huber:2008hg}. This fit is optimized for a frequency range close to the peak frequency, ${f}_0$, given by Eq.~\eqref{eq:freq}, and will be used in our numerical analysis for the prediction of the GW spectrum. 

Once the spectrum is known, we can also compute the characteristic amplitude produced by the stochastic gravitational waves around frequency $f$ given by~\cite{Moore:2014lga}
\begin{align}
h_c(f) \ & \ = \ \sqrt{\frac{3}{2\pi}} \: \frac{H_0\Omega_{\rm GW}}{f} \nonumber \\
\ & \ \simeq \ (1.3\times 10^{-18})\left[\Omega_{\rm GW}(f)h^2\right]^{1/2}\left(\frac{1~{\rm Hz}}{f}\right)\,.
\end{align}

\begin{figure}[t!]
\includegraphics[width=8cm]{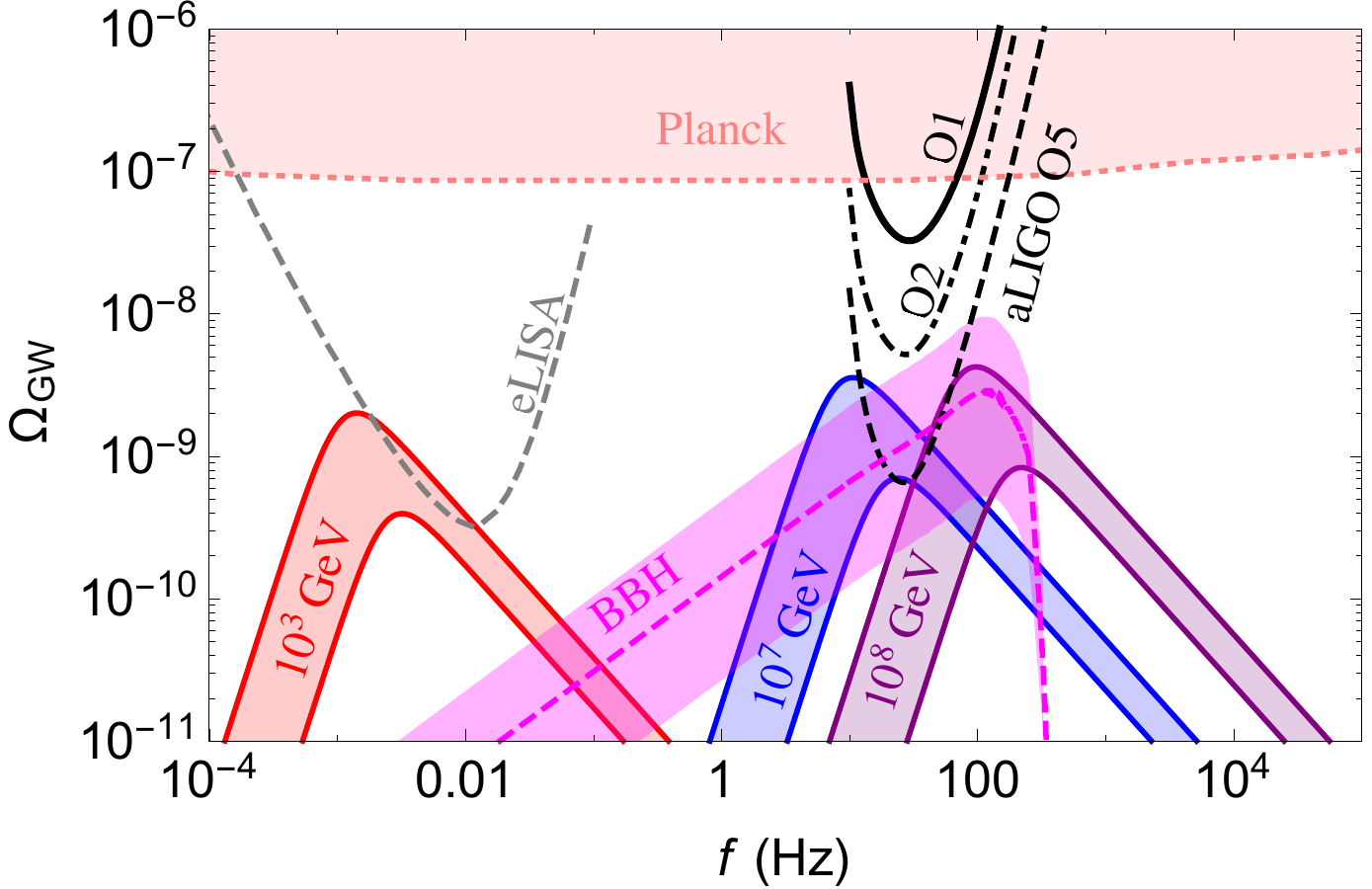}
\caption{Stochastic gravitational-wave spectrum from a first-order cosmological phase transition occurring at various critical temperatures (shown in GeV), with the variation of $\beta/H_*$ and $g_*$ shown by the solid bands. For 
comparison, we also show the expected stochastic background from binary black hole mergers (BBH), along with the 90\% C.L. statistical uncertainty~\cite{TheLIGOScientific:2016wyq}. The current 95\% C.L. upper limit from a recent cosmological data set (Planck)~\cite{Lasky:2015lej} is shown by the pink shaded region. The current advanced LIGO (aLIGO) sensitivity (O1), and the future observing run O2 (2016-17) and O5 (2020-22) sensitivities at $1\sigma$ C.L.~\cite{TheLIGOScientific:2016wyq} are shown by the black solid, dot-dashed and dashed curves, respectively. A projected sensitivity of eLISA~\cite{Caprini:2015zlo} is shown by the dashed gray curve.}
\label{fig:gw}
\end{figure}

In Figure~\ref{fig:gw}, we have shown the GW spectrum expected from a generic strong first-order phase transition as given by Eq.~\eqref{eq:gw2} for various representative values of the critical temperature $T_*=10^3,10^7,10^8$ GeV. Here, the band in each of the solid curves shows the uncertainty in the theoretical prediction due to variation in $\beta/H_*$ and $g_*$. The upper curves in each band show the optimistic case with $\beta/H_*$ estimated as in Eq.~\eqref{betaH} and with $g_*=106.75$ for the SM d.o.f., whereas the lower curves are for representative $\beta/H_*$ values twice of that given in Eq.~\eqref{betaH} and with $g_*=220$ for the MSSM d.o.f.. For larger values of $\beta/H_*\gtrsim 100$, the strength of the GW signal falls below the LIGO design  sensitivity. Similarly, adding more d.o.f. to the model will decrease the strength of the GW signal, since it gets more redshifted [cf.~Eq.~\eqref{eq:om}], and hence, is not favorable for its detectability. Note that 
if the frequency and energy density of the stochastic GW signal are determined with high precision in the future, one might be able to actually determine the value of $g_*(T_*)$ using Eqs.~\eqref{eq:freq} and \eqref{eq:om}, provided the signal spectrum exhibits the spectral features as predicted by a first-order phase transition [cf.~Eq.~\eqref{eq:gw2}] and a more accurate calculation of $\beta/H_\ast$ is available.

The Bayesian estimates of the current advanced LIGO detector sensitivity (O1) and the future observing run (O2 and O5) sensitivities at $1\sigma$ C.L. to generic power-law signals of stochastic gravitational wave~\cite{TheLIGOScientific:2016wyq} are also shown in Figure~\ref{fig:gw}. Comparing our results with these power-law integrated curves, we find that the first-order phase transitions occurring at critical temperatures around $10^7$--$10^8$ GeV can give rise to observable gravitational waves at LIGO/VIRGO detectors operating for 2 years at the design sensitivity in O5. On the other hand, the first-order phase transitions occurring at lower energies close to the EW scale can only be accessible to next-generation GW experiments, such as eLISA~\cite{AmaroSeoane:2012km}, which are sensitive to lower frequencies. As an illustration, we show in Figure~\ref{fig:gw} the projected sensitivity curve of a representative eLISA configuration C4~\cite{Caprini:2015zlo}. For the sensitivity of other future experiments in the low frequency range, see e.g.~\cite{Moore:2014lga}. 

For a possible distinction between the stochastic GW background discussed here and that due to unresolvable astrophysical sources, we also show in Figure~\ref{fig:gw} (magenta dashed curve) the expected stochastic background from
binary black hole mergers~\cite{Wu:2011ac}, as recently reevaluated in light of the GW150914 event~\cite{TheLIGOScientific:2016wyq}. The magenta shaded region around this curve shows the 90\% C.L. statistical uncertainty, propagated from the local rate measurement, on the total background~\cite{TheLIGOScientific:2016wyq}. Since this spectrum has a weaker power-law dependence, $\Omega_{\rm GW}\propto f^{2/3}$~\cite{Wu:2011ac}, a future world-wide network of more than two GW detectors, such as LIGO~\cite{TheLIGOScientific:2014jea}, VIRGO~\cite{TheVirgo:2014hva}, GEO600~\cite{Luck:2010rt}, KAGRA~\cite{Somiya:2011np}, and LIGO-India~\cite{Unnikrishnan:2013qwa}, can in principle separate this astrophysical signal from the one potentially arising due to a cosmological phase transition or from events taking place after cosmic inflation, and therefore, can provide a unique, powerful probe of the BSM physics at high scales. 

For completeness, let us also briefly discuss the primordial GW spectrum as predicted by inflation~\cite{Mazumdar:2010sa, Jinno:2011sw}. The simplest assumption for the resulting GW energy-density spectrum is a power-law: 
\begin{align}
P_t(f) \ = \ A_t\left(\frac{f}{f_{\rm CMB}}\right)^{n_t}, 
\label{power}
\end{align}
 where $f_{\rm CMB}=(1/2\pi)0.05~{\rm Mpc}^{-1}$, $n_t$ denotes the spectral index and $A_t$ is the amplitude of the primordial tensor perturbations, which is conventionally re-expressed in terms of the tensor-to-scalar ratio $r\equiv A_t/A_s$, where $A_s$ is the amplitude of the primordial power spectrum of (scalar) density perturbations. For the minimal inflation, using the consistency relation $n_t=-r/8$, we obtain a primordial GW spectrum of 
\begin{align}
\Omega_{\rm GW} \ = \ \frac{3}{128}\: r\: A_s\: \Omega_r \, ,
\end{align}   
where $\Omega_r$ is the total radiation energy-density evaluated today. Given the upper limit of $r<0.07$ at 95\% C.L. from a joint analysis of Planck and BICEP2/Keck array data~\cite{Array:2015xqh}, and the measured values of $A_s=(2.2\pm 0.1)\times 10^{-9}$ and $\Omega_r=2.473\times 10^{-5}h^{-2}$ from the Planck data~\cite{Ade:2015xua, Ade:2015lrj}, we obtain $\Omega_{\rm GW} \lsim 2\times 10^{-16}$ for the frequency range of interest in Figure~\ref{fig:gw}, which is far too small for LIGO sensitivity. 
Nevertheless, a significant blue tilt can lead to an enhanced power-spectrum~\cite{Kuroyanagi:2014nba, Lasky:2015lej}, but it has to be reanalyzed in light of new data from Planck and BICEP2/Keck array~\cite{Array:2015xqh}. Such a blue-tilted GW can in principle be obtained in certain early Universe cosmology models, which can provide an almost scale-invariant matter power spectrum but blue-tilted in primordial GW spectrum~\cite{Biswas:2014kva}. In any case, this primordial GW spectrum will have a weaker frequency dependence of $f^{n_t}$ with $n_t\lsim 0.36$ at 95\% C.L.~\cite{Lasky:2015lej}, which should be distinguishable from the astrophysical spectrum, as well as from that induced due to a first-order phase transition shown in Figure~\ref{fig:gw}. 

The current 95\% C.L. {\it integral} upper limit on $\Omega_{\rm GW}$ from a recent cosmological data set~\cite{Pagano:2015hma, Lasky:2015lej} is also shown in Figure~\ref{fig:gw} (pink shaded region labeled ``Planck"). Note that the conversion from the integral limit on $\int d(\ln f) \: \Omega(f)$ over a given range of frequencies to a limit on $\Omega(f)$ as we show in Figure~\ref{fig:gw} must assume a power-law spectrum with a known cut-off frequency, which we choose to be $f_{\rm max}=1$ GHz, corresponding to an energy scale of inflation $T=10^{17}$ GeV. This cosmological constraint rules out any inflationary models with a blue tilt $n_t>0.36$ at 95\% C.L. for $r=0.11$~\cite{Lasky:2015lej}, and with the latest constraint on $r<0.07$~\cite{Array:2015xqh}, the upper limit on $n_t$ is expected to be even (slightly) stronger, depending on the reheating temperature. 


\section{BSM Scenarios}\label{sec:4}

In this section, we point out the consequences for BSM physics which might potentially give rise to gravitational waves from first-order phase transition with a peak frequency around the LIGO sensitivity. Typically, the first-order phase transition can be mimicked by a scalar condensate with the following potential:
\begin{align}\label{scalar-pot}
V(\phi,\chi) \ = \ \frac{1}{4!}g^2\left(\phi^2-v_*^2\right)^2+ \frac{1}{2} h\phi^2\chi^2\,,
\end{align}
where $g,~h$ are coupling constants, $v_*$ is the vacuum expectation value (VEV) of the $\phi$ field responsible for the phase transition, and the
$\chi$ field belongs to the d.o.f in thermal bath.
If $\phi$ is a real scalar condensate, then we can avoid the domain-wall formation~\cite{Kibble:1976sj}.  
For $v_*\ll 10^{15}$~GeV, cosmic strings associated to the phase 
transition are also harmless~\cite{Hertzberg:2008wr}. Typically, at high temperatures, $\chi$ would induce thermal correction to the $\phi$ field proportional to $T^2\phi^2$ potential around 
$\phi=0$,
\begin{align}
V_T(\phi) \ = \ \frac{1}{24}h(T^2-T_\ast^2)\phi^2 +\cdots \, ,
\label{thermal}
\end{align}
where 
\begin{align}
T_\ast \ = \ \sqrt{\frac{2g}{h}} \: v_* 
\end{align}
is the critical temperature. For temperatures well above the critical temperature, $T \gg T_\ast$, the potential is in a symmetric phase, but as the temperature decreases,  
the negative mass-squared term in the zero-temperature scalar potential given by Eq.~(\ref{scalar-pot}) wins over the thermal mass term in Eq.~\eqref{thermal}, and the phase transition occurs.  

The first-order phase transition occurs when $T_\ast/ v_\ast \leq 1$.
Now, in order to obtain the frequency range and $\Omega_{\rm GW}$ accessible at LIGO, we expect the associated new physics must be in the vicinity of the required critical temperature of 
$T_\ast \sim 10^7$--$10^8$ GeV, as shown in Figure~\ref{fig:gw}. A pertinent question is what could be the interesting possibilities relevant for BSM physics which might occur at such high scale. Here we will address this question in a general, qualitative way, without going into the gory details of the model building  aspects, which are postponed to a future work. Also, our list of examples is by no means exhaustive and there are other possibilities for an observable gravitational wave from some particle physics processes not mentioned here; see e.g.~\cite{Jaeckel:2016jlh} for a recent discussion of detectable gravitational waves from a dark sector.

\subsection{Peccei-Quinn Symmetry}\label{sec:4a}

The first example we consider here is the high-scale breaking of a $U(1)_{\rm PQ}$  symmetry~\cite{Peccei:1977hh}. In this case, 
we have to assume the scalar field $\phi$ in Eq.~\eqref{scalar-pot} to be complex, so that the pseudo-scalar axion belongs to its imaginary component to explain the smallness of the QCD $\theta$-parameter. The part of the scalar potential responsible for phase transition is of the ``wine bottle" form   
\begin{align}
V(\phi) \ = \ g\left(|\phi|^2-\frac{f_a^2}{2}\right)^2 \, ,
\end{align}
where $f_a$ is known as the axion decay constant. In this case, we require the PQ symmetry breaking scale, synonymous with $f_a$, to be close to $T_\ast \sim 10^{7}-10^8$~GeV range in order to give an observable GW signal in the LIGO frequency range. 
Now considering the finite temperature effects for the PQ scalar $\phi$, a strong first-order phase transition can happen if the PQ field couples to some fields in the thermal bath, e.g. to the SM Higgs via 
quartic coupling, i.e. $h|\phi|^2|H|^2$, and if the SM Higgs fields are in thermal equilibrium after the epoch of reheating. Naturally we will have to assume a scenario where the reheating temperature of the Universe is larger than $f_a$. In the region where temperature corrections to the $\phi$ mass become important, it is possible for a first order phase transition to occur when $m^2_\phi(T) < 0$ below $T< T_\ast$, see Eq.~(\ref{thermal}). In this scenario, one can estimate $\beta/H_\ast$ in a thin-wall limit, as in Eq.~\eqref{betaH2},  which would yield ${\beta}/{H_\ast}\simeq 10^{3}g\sim {\cal O}(10-100)$ for $g\sim 0.01-0.1$.

The PQ breaking scale of $f_a\sim 10^7-10^8$ GeV is still allowed by the current experimental constraints on $f_a$; see e.g.~\cite{Hertzberg:2008wr}. A special note should be made 
to the searches for axion-like particles $a$ produced in the decay $B^0\to K^{*0}a$, with $K^{*0}\to K^+\pi^-$ and $a\to \ell^+\ell^-$, which impose stringent constraints on $f_a$ in the multi-TeV range~\cite{Batell:2009jf, Aaij:2015tna}. Future dedicated searches at LHCb and in $B$-factories can in principle access the range of $f_a$ that gives rise to an observable signal in LIGO/VIRGO. 

The axion being nearly massless during inflation can also give rise to axion iso-curvature perturbations~\cite{Turner:1990uz}. However, for $f_a\sim 10^{7}$--$10^8$~GeV, the iso-curvature perturbations created during inflation for $H_{\rm inf}\leq f_a$ are negligible and well within the current 
Planck limits~\cite{Ade:2015lrj}, depending of course on the initial misalignment angle $\theta\sim a/f_a$~\cite{Harigaya:2015hha}, where $a$ is the QCD-type axion. One challenge which may arise in this case is the domain wall problem, and the associated constraints~\cite{Hiramatsu:2013qaa}. However, the domain walls may not be created if the initial fluctuations in the axions after inflationary phase do not restore the symmetry via parametric resonance. The latter part is a model-dependent issue, which mainly depends on how the inflaton field responsible for reheating the Universe couples to the PQ field.

\subsection{High-scale Supersymmetry} \label{sec:4b}
Now let us consider a few examples in the SUSY context. Although a weak-scale SUSY is highly attractive due to its ability to solve the hierarchy problem in a natural manner, the lack of its evidence in the current LHC data suggests that the SUSY-breaking scale could be higher, and therefore, it is important to explore other opportunity windows, such as the one proposed here, to indirectly probe this scale.  

{\bf NMSSM}: The minimal supersymmetric version of the SM, namely, the MSSM, has one dimensionful parameter in the superpotential, known 
as the $\mu$-term, i.e., $\mu H_uH_d$, where the VEVs of the $SU(2)$ doublets $H_u$ and $H_d$ give masses to up-type and down-type quarks,  respectively. In order to address the hierarchy problem, 
one requires $\mu \sim {\cal O}({\rm TeV})$. A simple solution within the context of the so-called NMSSM scenario~\cite{Ellwanger:2009dp} is to extend the MSSM field content by an additional singlet chiral 
superfield $S$ which, after getting a VEV, dynamically generates the $\mu$-term. One can in fact start with a discrete $Z_3$-symmetry being imposed on the singlet $S$, with a superpotential
\begin{align}
W \ = \ W_{\rm MSSM} + \lambda SH_uH_d+\kappa S^3 \, ,
\end{align}
where $W_{\rm MSSM}$ represents the standard interactions between Higgs doublets and quarks/leptons in the MSSM, and $\lambda$ and $\kappa$ are dimensionless 
couplings. In this case, $\mu=\lambda \langle S\rangle $, and it is certainly possible to have the singlet VEV $\langle S\rangle \sim 10^{7}$--$10^8$ GeV and $\lambda \sim 10^{-4}$--$10^{-5}$, while being consistent with all existing constraints. In this respect, 
we are contemplating two phase transitions, i.e. one at high scale due to the singlet VEV and another at the electroweak scale due to the VEVs of $H_{u,d}$. One has to do a numerical scan of the NMSSM parameter space to find the viable region where this happens, but this is beyond the scope of this paper. For an observable GW signal at LIGO, all that we require is that the high scale
phase transition should be of first order. In the parameter region where thermal corrections to the bare singlet mass become important, it is possible for a first order phase transition to occur when $m^2_{S}(T)< 0$. Since the thermal contribution to the singlet mass would be proportional to $\sim (\lambda^2+\kappa^2)T^2S^2$, one can imagine obtaining a first-order phase transition with $\beta/H_\ast\sim {\cal O}(10-100)$ for a wide range of $\lambda$ and $\kappa$ in the thin-wall limit.

 With such high-scale SUSY breaking, a sizable radiative correction to the singlino mass is possible, which enlarges the singlino dark matter parameter space~\cite{Ishikawa:2014owa}.
The impact of high-scale SUSY breaking on the GW spectrum has been discussed in Ref.~\cite{Watanabe:2006qe}. The physics of first-order phase transition in these models will be very similar to the EW-scale NMSSM~\cite{Menon:2004wv}, but now we will have to imagine the phase transition occurring at VEVs close to $v_* \sim 10^{7}$--$10^8$~GeV, if we were to constrain this scenario from LIGO data.

{\bf Split-SUSY}: Another scenario where the SUSY breaking scale could naturally be around $10^7$--$10^8$ GeV is split-SUSY~\cite{Giudice:2004tc}. In fact, for the observed value of the Higgs mass around 125 GeV, such SUSY breaking scales lead to a stable or metastable vacuum~\cite{Giudice:2011cg}. Moreover, as shown recently in Ref.~\cite{Evans:2016htp}, there exist SUSY versions of the relaxion mechanism~\cite{Graham:2015cka} that naturalize such high-scale SUSY models, while preserving the QCD axion solution to the strong CP problem.  If the SUSY breaking sector undergoes a first-order phase transition, such as in Ref.~\cite{Gies:2009az} for instance, this would be an ideal source for generating gravitational waves potentially testable by the future GW detector network. 

In the simplest framework, one can imagine that a hidden sector field $\phi$ undergoes a first-order phase transition at the relevant scale of 
$10^7$--$10^8$ GeV, while the finite VEV of $\langle \phi \rangle \sim v$ is responsible for dynamical SUSY-breaking in the hidden sector. This can be communicated to the visible sector by a messenger field. 
As a concrete example, we mention the $Z'$-mediated SUSY breaking~\cite{Langacker:2007ac}, where both visible and hidden sector fields are charged under $U(1)'$. In this case, the scalar components of the chiral superfields acquire large masses close to the SUSY breaking $Z'$-ino mass at one-loop level, whereas the MSSM gauginos get much smaller masses at two-loop level. Following the previous discussions, one can then include finite-temperature corrections to the effective potential, and in principle, obtain a reasonable $\beta/H_\ast\sim {\cal O}(10-100)$ for observable $\Omega_{\rm GW}\sim 10^{-9}$ due to enough freedom in the parameter space. 

{\bf Sneutrino-Mediation}: As far as the mediator of SUSY breaking is concerned, our discussion is generically applicable irrespective of the particulars of the SUSY breaking mechanism, since it only relies on the requirement that the symmetry breaking process must be a first-order phase transition. As mentioned above, the SUSY breaking could occur due to some hidden sector dynamics. Alternatively, one can envisage SUSY-breaking in the visible sector, e.g. by the VEV of a right-handed sneutrino~\cite{Mohapatra:1986uf}, at a scale around $10^{7}-10^8$ GeV. A first-order phase transition can be induced by finite-temperature corrections similar to Eq.~(\ref{thermal}), but now the relevant couplings are  $h$ and  $\kappa$ of  the superpotential term $W\supset hNH_uL +\kappa N^3$, where $N$ and $L$ are the right-handed neutrino and lepton superfields, respectively. As in the NMSSM case, the required values of $\beta/H_\ast\simeq {\cal O}(10-100)$ can be obtained for a range of combinations of $h$ and $\kappa$ values.

 One advantage of this scenario is that via the $NH_uL$ term in the superpotential,
 one can induce appropriate soft terms and  provide a successful thermal/non-thermal leptogenesis mechanism~\cite{Ellis:2003sq}, while simultaneously explaining the neutrino masses within the usual seesaw framework~\cite{seesaw}. 
Such a low seesaw scale (as compared to the GUT scale) can be motivated from naturalness arguments~\cite{Vissani:1997ys}. Inflation and dark matter issues can also be addressed within this common framework~\cite{Allahverdi:2007wt}.  

Finally, we would like to remark that the first-order phase transition naturally satisfies one of Sakharov's conditions~\cite{Sakharov:1967dj} for dynamically generating the matter-antimatter asymmetry in the Universe~\cite{Rubakov:1996vz}. Therefore, a new window of opportunity to constrain high-scale baryogenesis scenarios can be opened up by the GW detectors following the stupendous success of LIGO, and the details of the correlation between the GW production and baryogenesis in a concrete theoretical framework is worth pursuing in the future.

\section{Conclusion}\label{sec:5}

In light of the recent direct detection of gravitational waves~\cite{Abbott:2016blz}, we have discussed the possibility of probing some beyond the Standard Model scenarios which could lead to a stochastic GW background of cosmological origin within the projected sensitivity reach of the advanced LIGO/VIRGO. One of the key features to exploit is the energy spectrum of gravitational waves, which can discriminate the stochastic background due to unresolved astrophysical sources   from those of cosmological origin, such as cosmological phase transitions and primordial inflation. 

We have mainly focused on the physical scenario of a first-order phase transition in a vacuum-dominated epoch, which can optimize the peak frequency and the corresponding peak fraction of energy density released in the gravitational waves to be within the LIGO sensitivity range, provided the scale of phase transition is around $10^7$--$10^8$ GeV. It is possible to conceive this first-order phase transition in the early Universe arising from a PQ symmetry breaking, with an axion decay 
constant, $f_a\sim 10^{7}$--$10^8$~GeV. Such a phase transition temperature could also point toward a high-scale SUSY breaking scenario
within MSSM and beyond, as well as naturally in the context of split-SUSY. A number of BSM physics issues can be addressed in these scenarios, such as the strong CP problem in the $U(1)_{\rm PQ}$ case, and baryogenesis, neutrino masses, origin of dark matter, and possibly, the scale of inflation in the SUSY case. 

To conclude, we believe the positive detection of gravitational waves by LIGO is the beginning of a new era not just for astrophysics, but also for cosmology as well as BSM physics. In particular, it provides an unprecedented opportunity to constrain various BSM physics scenarios at high energy scales not directly accessible by laboratory experiments. The precision GW astronomy promised by the world-wide network of GW detectors can make this dream a reality in the not-so-distant future. 

\section*{Acknowledgments}
A.M. would like to thank Alex Koshelev and Tomo Takahashi for helpful discussions. We also thank Jose No, Pedro Schwaller and Michael Williams for useful comments on the draft. The work of B.D. is supported by the DFG grant No. RO 2516/5-1. The work of A.M. is supported in part by the Lancaster-Manchester-Sheffield Consortium for Fundamental Physics  under  STFC  grant No. ST/L000520/1. 

\end{document}